# Raman spectra of twisted bilayer graphene close to the magic angle

Tiago C. Barbosa[1,2,τ], Andreij C. Gadelha[1,3, τ], Douglas A. A. Ohlberg[4], Kenji Watanabe[5], Takashi Taniguchi[6], Gilberto Medeiros-Ribeiro[7], Ado Jorio[1,ζ], Leonardo C. Campos[1,2, ζ,*]

[1] *Physics Department, Universidade Federal de Minas Gerais, Belo Horizonte, MG 31270-901, Brazil.*

[2] *Center of Technology in Nanomaterials and Graphene, Universidade Federal de Minas Gerais, Technological Park of Belo Horizonte, Belo Horizonte, MG, 31270-901, Brazil.*

[3] *Department of Physics, Department of Chemistry, and JILA, University of Colorado at Boulder, Boulder, Colorado 80309, USA.*

[4] *Microscopy Center, Universidade Federal de Minas Gerais, Belo Horizonte, MG 31270-901, Brazil.*

[5] *Research Center for Functional Materials, National Institute for Materials Science (NIMS), 1-1 Namiki, Tsukuba 305-0044, Japan.*

[6] *International Center for Materials Nanoarchitectonics, National Institute for Materials Science (NIMS), 1-1 Namiki, Tsukuba 305-0044, Japan.*

[7] *Computer Science Department, Universidade Federal de Minas Gerais, Belo Horizonte, MG 31270-901, Brazil.*

[τ] These authors contributed equally to this work.

[ζ] These authors equally supervised the project.

* Corresponding author. Tel.: 55 31 3409-6641. E-mail: lccampos@fisica.ufmg.br (Leonardo C. Campos).

ABSTRACT

**In this work, we study the Raman spectra of twisted bilayer graphene samples as a function of their twist-angles ($\theta$), ranging from 0.03º to 3.40º, where local $\theta$ are determined by analysis of their associated moiré superlattices, as imaged by scanning microwave impedance microscopy. Three standard excitation laser lines are used (457, 532, and 633 nm wavelengths), and the main Raman active graphene bands (G and 2D) are considered. Our results reveal that electron-phonon interaction influences the G band's linewidth close to the magic angle regardless of laser excitation wavelength. Also, the 2D band lineshape in the $\theta < 1º$ regime is dictated by crystal lattice and depends on both the Bernal (AB and BA) stacking bilayer graphene and strain soliton regions (SP) [1]. We propose a geometrical model to explain the 2D lineshape variations, and from it, we estimate the SP width when moving towards the magic angle.**

1. INTRODUCTION

Van der Walls heterostructures (vdW-H) are formed by assembling different 2D materials, and they may present distinct electrical, optical, and mechanical properties from their counterparts [2–4]. The study of vdW-H revealed that the properties of the final structure depend on the chosen materials and, in the case of twisted bilayer or few-layer materials, the relative twisting angle ($\theta$) between each atomic layer plays an important role [5]. In 2018, researchers showed that twisted bilayer graphene (TBG) devices with relative twist angles close to 1.1º, known as the magic angle ($\theta_M$) [6,7], behave as strongly interacting systems that allow unconventional superconductivity [8]. Such TBG devices present phase transitions from correlated insulators, conductors, superconductors, and even ferromagnetic phases [9–14]. This unique system also presents interesting electron-phonon related phenomena [15,16], which may shed light into the role of this interaction in the aforementioned phase transitions [17–20].

Raman spectroscopy has been widely used to study graphene devices due to its several advantages: it is a fast and non-destructive technique that allows investigation of strain, doping, disorder, electron-phonon interaction, and many other aspects of such devices [21–24]. Raman spectra of what we call here large-angle twisted bilayer graphene devices, which means $\theta$ ranging from 3º to 30º, has been largely studied [25,26,35,36,27–34] and a review on the topic can be found in Ref. [15]. The most important aspect in this large angle range is a plethora of new peaks originated from the moiré superlattice and an increase in the main G and 2D peaks intensity in certain twist angles, which in turn provides a simple way of accurately determining the TBG twist angle [31,33]. More recently, nano-Raman measurements revealed that TBG with angles smaller than $\theta_M$ undergo a self-organized crystal lattice reconstruction [1]. In this range of twist angles, TBG show periodical triangular areas of alternating Bernal (AB and BA) stacking domains, separated by shear soliton regions (SP), with AA-stacked regions at the vertices of the triangular areas [1].

This work presents micro-Raman study of TBG samples in the twist-angle range around the magic angle, i.e. for $\theta$ ranging from 0.03º to 3.40º. We investigated the Raman spectra of twisted bilayer graphene devices with three different excitation lasers, and we show that samples with angles close to 0º have their Raman spectra dominated by Bernal stacking bilayer graphene. Our data also reveal an increasing contribution of AA-stacking and soliton regions to the behavior of the sample with increasing angle towards the magic angle, as proposed theoretically in band structure calculations by Nguyen et al. [37]. Furthermore, we analyzed qualitatively electron-phonon interactions in TBG devices in terms of the Kohn anomaly process, and we related the increase in the Raman G band full width at half maximum close to the magic angle with the appearance of a flat band in the band structure of such systems [6,7,38].

2. METHODS

The graphene samples were obtained by the standard scotch tape method [39] and deposited over Si substrate covered by 285 nm thick thermally oxidized silicon. The TBG samples were fabricated through a tear-and-stack method [40] with the aid of a polydimethylsiloxene (PDMS) pyramid stamp reported elsewhere [1,41]. Described briefly, a polycarbonate membrane is attached to a PDMS pyramid stamp (PS), aligned over a region of the Si substrate containing the graphene flake, and pressed into contact with it while raising the sample temperature. A few minutes later, the sample is allowed to cool down, and the stamp is retracted from the substrate, causing the portion of flake adhering to the stamp to tear loose from the rest of the flake that remains on the substrate. After that, the substrate with the remaining graphene sheet is rotated to the desired twist angle relative to the fixed stamp orientation, and the whole process with the PS is repeated to pick up the rest of flake off the substrate. At last, the TBG is attached to the stamp and can be deposited over any chosen substrate. Additionally, we have picked up a hexagonal boron nitride (hBN) flake as a supporting substrate for the TBG, an atomically flat surface free of dangling bonds or charge traps [42]. In the end, we deposited our heterostructure over a microscope glass slide, as depicted in Fig. 1(a). The fabricated TBG samples were then characterized by scanning microwave impedance microscopy (sMIM), revealing different moiré patterns [43] for different twist angles as shown in Fig. 1 for (b) $\theta = 0.17°$, (c) 1.20°, and (d) 3.40°. Each twist angle was estimated by the expression $\theta = 2.\arcsin(a_G/2.L_M)$ [44], where $a_G = 0.246$ nm is the graphene crystal lattice constant and $L_M$ is the $\theta$-dependent moiré superlattice periodicity, obtained from the sMIM images.

Thereafter, we acquired hyperspectral Raman maps of the TBG using a WITec Alpha300R system, with 1.0 mW laser power at the sample, focused by a 100x objective with NA = 0.9. Finally, hyperspectral Raman maps and sMIM images are combined, enabling identification of homogeneous moiré patterns from which the Raman data were acquired. For further information on the Raman maps and sMIM images combination, please refer to the Supporting Information. This process was performed using three different excitation laser wavelengths, namely 457 nm, 532 nm and 633 nm.

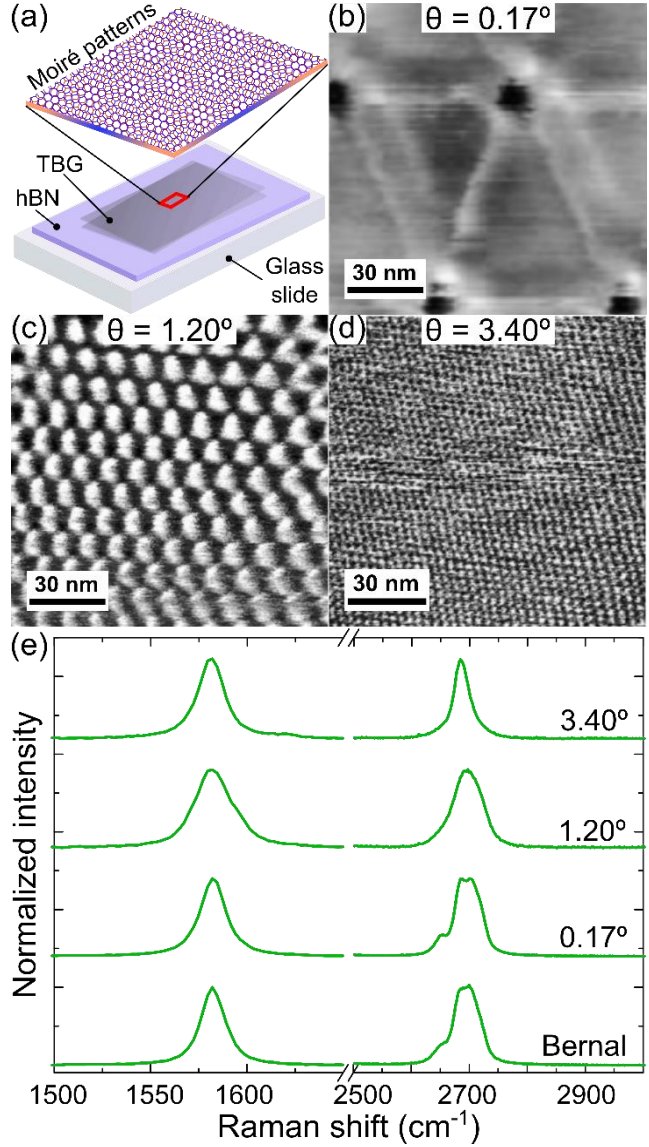

Fig. 1: (a) Schematics of our hBN-TBG sample produced through a tear-and-stack method with the aid of a PDMS pyramid stamp deposited over a glass slide – inset shows a schematic of a moiré pattern that can be found in TBG samples. Superlattices for samples with (b) $\theta$ = 0.17º, (c) 1.20º, and (d) 3.40º twist angles, measured by sMIM. (e) G and 2D band Raman spectra of a Bernal bilayer graphene and the samples shown in (b), (c), and (d) (see labels near each spectrum), acquired with a 532 nm excitation laser.

## 3. GENERAL TRENDS OF THE RAMAN SPECTRUM

The first comparison between Raman spectra of Bernal stacking bilayer graphene and TBG with different twist angles is depicted in Fig. 1(e), measured with a 532 nm excitation laser wavelength. We focus our attention on the most prominent graphene

Raman peaks: the G band around 1580 cm$^{-1}$, which is a first-order Raman peak associated with the doubly degenerate (iTO and LO) phonon mode at the Brillouin zone center, and the 2D band around 2700 cm$^{-1}$, which is a dispersive second-order Raman peak involving two iTO phonons near the *K* point at the edges of the Brillouin zone [22,45,46]. At first glance, the major difference in the G band with respect to $\theta$ is an increase in its linewidth ($\Gamma_G$) at the angle close to the magic angle (see $\theta = 1.20°$ in Fig. 1(e)) [1]. Regarding the 2D band, note that the $\theta = 0.17°$ spectra resemble the bilayer graphene Bernal stacking spectrum [45], whereas the spectra for $\theta = 1.20°$ and 3.40° reveal different shapes.

To further investigate the Raman spectra dependence on the TBG twist angle, several samples ranging from 0.03° to 3.40° were measured with different excitation laser wavelengths, as shown in Fig. 2 for (a) 457 nm, (b) 532 nm, and (c) 633 nm – the data are normalized with respect to each G and 2D bands spectra for clarity. The G band of each spectrum is fitted with a single Lorentzian peak and analyzed in terms of its frequency ($\omega_G$) and full width at half maximum ($\Gamma_G$). There is no clear dependence of $\omega_G$ on the twist angle, as highlighted in Fig 3(a). On the other hand, there seems to be a peak broadening for samples with $\theta$ close to $\theta_M$ – this trend will be further discussed in more detail. The 2D band of each spectrum is evaluated in terms of its line shape and of a medium frequency ($\omega_{2D}^{\sim}$), which was defined as $\omega_{2D}^{\sim} = \int(I(\omega).\omega)d\omega / \int I(\omega).d\omega$, where $I(\omega)$ is the Raman spectrum intensity, $\omega$ is its frequency, and the integrals were evaluated in the 2D band region (2600 cm$^{-1}$ < $\omega$ < 3000 cm$^{-1}$ range). Yet again, there is no clear dependence of $\omega_{2D}^{\sim}$ on $\theta$, while preserving the expected dispersive behavior with respect to the energy of the incident laser (Fig. 3(b)) [22,45]. Note that the 2D band line shape of samples with twist angles close to 0° resembles bilayer graphene with Bernal stacking, see Fig. 2, which can be described in terms of four Lorentzian peaks [22,45]. In Fig. 2, we can also see that the 2D band line shapes gradually change and become more and more similar to AA stacking or SP soliton regions (Fig. 3(b) in Ref. [1]) with increasing $\theta$ towards the magic angle, which can be more apparent for $\theta \geq 0.70°$. This tendency is related to a more significant contribution from AA stacking or soliton regions to the measured Raman spectra in samples with $\theta \geq 0.70°$, as noted by Nguyen et al. [37], which will be discussed in more detail in the next section.

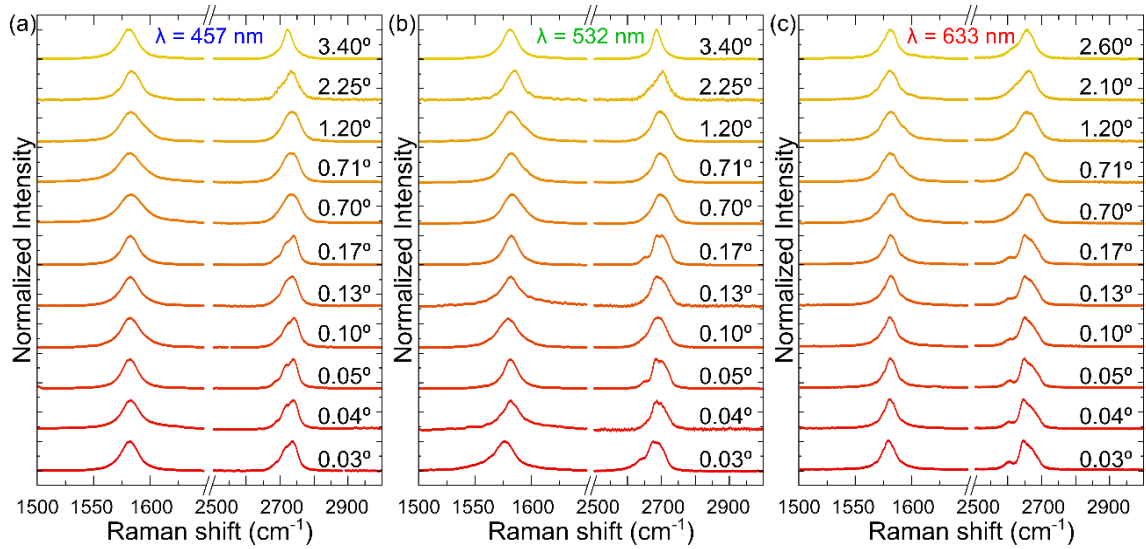

Fig. 2: Raman spectra of TBG with twist angles ranging from 0.03º to 3.40º (see labels near each spectrum), measured with (a) 457 nm, (b) 532 nm, and (c) 633 nm excitation lasers wavelengths.

The use of Raman spectroscopy to evaluate mechanical strain and charge doping in graphene was introduced by Lee et al. [47]. The correlation between the medium 2D band and G band frequencies (Fig. 3(c)) reveals that the data spread mainly at a 2.2 slope, indicating that our samples present different mechanical strain and no doping differences among samples [47]. Despite revealing different mechanical strain, there is no clear dependence of strain on $\theta$.

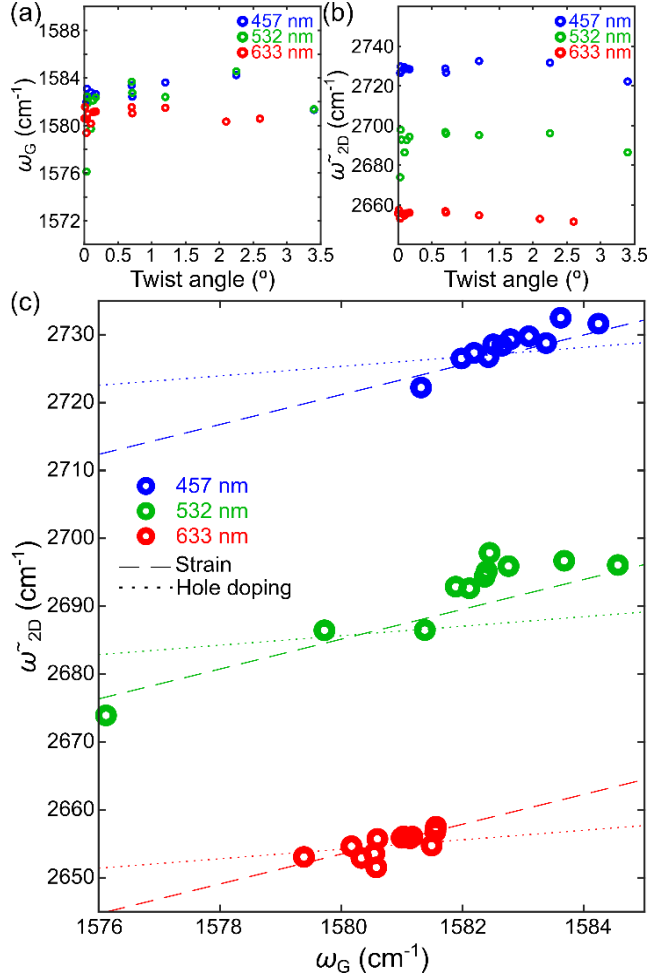

Fig. 3: Raman shift of (a) G band, and (b) medium 2D band. (c) Medium 2D band Raman shift ($\omega_{\widetilde{2D}}$) and G band Raman shift ($\omega_G$) relation – the dashed (dotted) lines with 2.2 (0.7) slope reveal the strain (doping) tendencies [47], rigidly shifted to match the data for each excitation laser.

4. THE 2D BAND LINE SHAPE

The 2D band (also called G´ in the literature) is the result of a combination of allowed double-resonant two-phonon Raman scattering processes, with the lineshape depending on the specific electronic and phononic structure of the material, differing for different types of $sp^2$ carbons [45,48–53], as well as on the twist-angle in the large-angle regime [27]. As mentioned before, in the angle regime studied here, there is a gradual change in the 2D band line shape with increasing twist angle, which can be noticed in Fig. 2 for all laser lines. Starting from $\theta = 0$, each spectrum can be fitted with a combination of a Bernal (AB and BA) 2D band spectrum ($S_{AB}$) and a free pseudo-Voigt function ($S_{pV}$) [54]. $S_{AB}$ is a 4-peaks structure which depends on the excitation laser line [45,55]. $S_{pV}$ is a generic function that varies from a Lorentzian to a Gaussian distribution

[56,57], used here as a practical approximation to address the emission from the AA/SP regions, which are present in the reconstructed TBG in low twist angles [1]. Therefore, the total spectrum ($S_T$) is a combination of these two features ($S_T = \alpha S_{AB} + \beta S_{pV}$), with α and β measuring the relative contribution from $S_{AB}$ and $S_{pV}$, respectively, i.e. (α → 1, β → 0) for $\theta \to 0$ and (α → 0, β → 1) in the large-angle limit. Indeed, at twist angles larger than $\theta_M$, the $S_{pV}$ function alone fits the 2D bands considerably well.

The θ-dependent 2D spectral change is explained as follows: at angles close to 0° (Fig. 4(a)), the spectrum is dominated by Bernal regions, and the contribution of the pseudo-Voigt function is insignificant (α ≈ 1, β ≈ 0). Closer to the middle range between 0° and the magic angle (Fig. 4(b)), we see an increase in the contribution of the pseudo-Voigt function, indicating a more considerable contribution from AA/SP regions to the overall spectrum. Closer to the magic angle (Fig. 4(c)), the contribution of the pseudo-Voigt function is stronger than Bernal's one indicating a dominance of the AA/SP region or from the large-angle TBG spectra. Finally, at twist angles larger than $\theta_M$ (Fig. 4(d)), insignificant contribution from the Bernal spectrum is observed, and the pseudo-Voigt function entirely dominates (α ≈ 0, β ≈ 1), as expected, since the Bernal stacking ceases to exist above $\theta_M$ [58].

In order to relate this general trend of the 2D band spectra with the TBG lattice, a geometric model based on the size of the AB/BA and AA/SP regions was used. The contribution of each component to the overall spectrum area was analyzed, and related to the α and β spectral contributions. Then, these parameters can be related with their geometric counterpart in the TBG lattice, which were called $A_{AB}$ for the Bernal area, and $A_{SP}$ for the AA/SP area (inset in Fig. 4(e)). At last, the α/β ratio is given by the $A_{AB}/A_{SP}$ ratio, and can be expressed in terms of the soliton width ($W_{SP}$), the TBG twist angle (θ) and the graphene crystal lattice constant ($a_G$):

$$\frac{\alpha}{\beta} = \frac{A_{AB}}{A_{SP}} = \frac{\dfrac{a_G}{4.\sin\dfrac{\theta}{2}} - W_{SP}.\sqrt{3} + \dfrac{3.W_{SP}^2.\sin\dfrac{\theta}{2}}{a_G}}{W_{SP}.\sqrt{3} - \dfrac{3.W_{SP}^2.\sin\dfrac{\theta}{2}}{a_G}}$$

(eq 1)

Therefore, the above expression was used to fit the α/β ratio as a function of the twist angle, with $W_{SP}$ as a free adjusting parameter, as shown in Fig. 4(e) in a log-log scale for all laser lines. After the analysis of the data with the proposed geometric model, the obtained soliton width was approximately $W_{SP} = (4 \pm 3)$ nm, which agrees quite well with theory [59] and measurements performed through different experimental techniques [60,61]. In Fig. 4(e) we exhibit the best fitted $\alpha/\beta$ curve with $W_{SP} = 4$ nm (dashed black curve), but we also provide $W_{SP} = 2$ nm (blue) and $W_{SP} = 7$ nm (orange) curves for comparison. The tendency observed in Fig. 4(e) is not laser-dependent which corroborates the proposed geometric model. At this point, it is important to highlight that our model ignores a possible $W_{SP}$ dependence on the twist angle [59] and, despite being a simple phenomenological model, it describes the phenomenology and our data significantly well, introducing a new approach to address the width of solitons in the reconstruction regime of twisted bilayer graphene samples. The $W_{SP}$ value found here should be applicable to the soliton width towards small moiré pattern unit cells, where the the soliton width plays a more relevant role for the $A_{AB}/A_{SP}$ ratio.

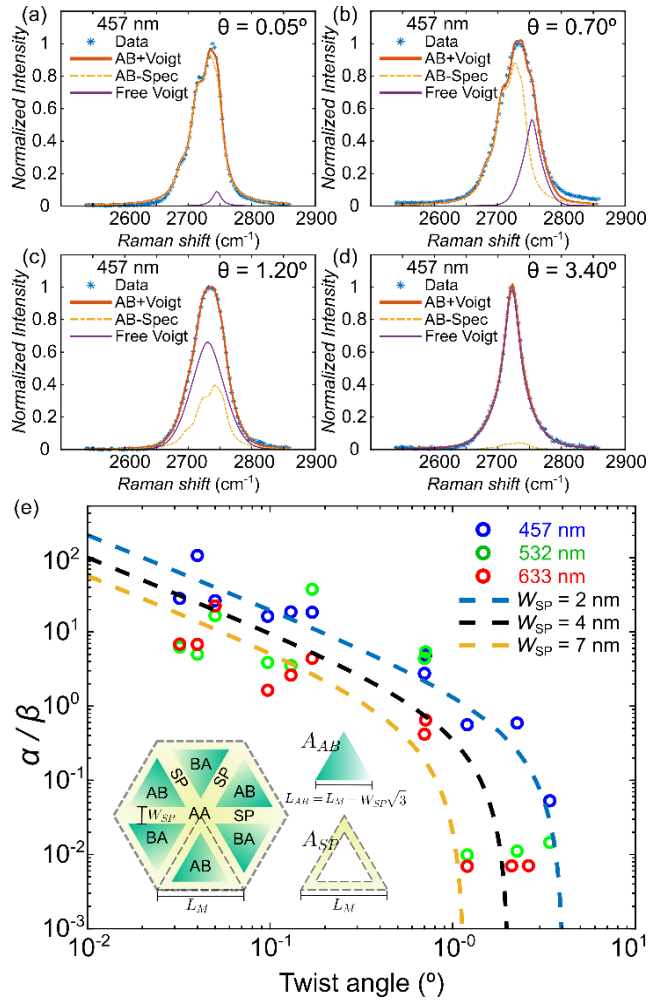

Fig. 4: 2D band spectra for samples with $\theta$ equals to (a) 0.05°, (b) 0.70°, (c) 1.20° and (d) 3.40°, measured with a 457 nm laser wavelength and fitted with a combination of a Bernal

stacking spectrum, $S_{AB}$, and a pseudo-Voigt function, $S_{pV}$. The α and β fitting parameters were left free to vary, without any bonding condition; (e) α/β ratio as a function of twist angle in a log-log scale for all laser lines, with the data points coming from the spectral analysis, and the black dashed-line representing the data points fitted with the geometric model summarized in the inset (see text for details) for $W_{SP}$ = 4 nm – curves for $W_{SP}$ = 2 nm (blue dashed-line) and 7 nm (orange dashed-line) were also plotted for comparison.

## 5. THE G BAND FWHM

Now we turn our attention to the dependence of $\Gamma_G$ on $\theta$. Previous works reported an increase in $\Gamma_G$ with decreasing twist angle in the large-angle range below $\theta = 5°$ [36]. Our results reveal a tendency to reach a maximum peak width close to the magic angle for all measured laser lines (Fig. 5(a), (b), and (c) for 457 nm, 532 nm, and 633 nm, respectively). Fig. 5(c) was adapted from Gadelha et al. [1], where a single excitation energy (633 nm) was used.

The electron-phonon interaction process in graphene is known to cause an increase in the G band full width at half maximum [55,62–65]. In this process, an initial phonon with frequency $\omega_q$ excites an electron-hole pair, which recombines emitting a phonon with energy $E$, as depicted in Fig. 5(d) [55,62–65]. The electron-hole pair creation rate and the recombination rate depend on the allowed electronic states involved and the phonon range of energy. So, when the Fermi energy is in the charge neutrality point, there is a large probability of electron-phonon interaction, thus decreasing the phonon lifetime, as stated by the uncertainty principle. This process reveals itself as an increase in the G Raman peak width [55,62–65]. In TBG, twist angle changes the energy band and electronic density of states. In this sense, in TBG, the electron-phonon interaction turns stronger due to an increase in the electronic density of states (DOS). Fig. 5 also presents three major cases: $\theta < \theta_M$, $\theta = \theta_M$ and $\theta > \theta_M$, as for (e) $\theta = 0.74°$, (f) $\theta = 1.08°$ (magic angle), and (g) $\theta = 3.15°$ (adapted from Nguyen et al. [37]). Despite the complex band structure of the $\theta = 0.74°$ case, it is clear that its DOS remains in the same order of magnitude as the DOS for $\theta = 3.15°$. Nevertheless, the band structure of the $\theta = 1.08°$ reveals a flat band at zero energy which causes the DOS to escalate. Therefore, it is expected an increase in $\Gamma_G$ close to the magic angle, regardless of the laser excitation energy, as revealed by Fig. 5 and Gadelha et al. [1]. Nonetheless, it is important to highlight that the presented explanation is qualitative. Further experiments and theoretical studies, ideally with doping, need to be performed to describe the data results quantitatively.

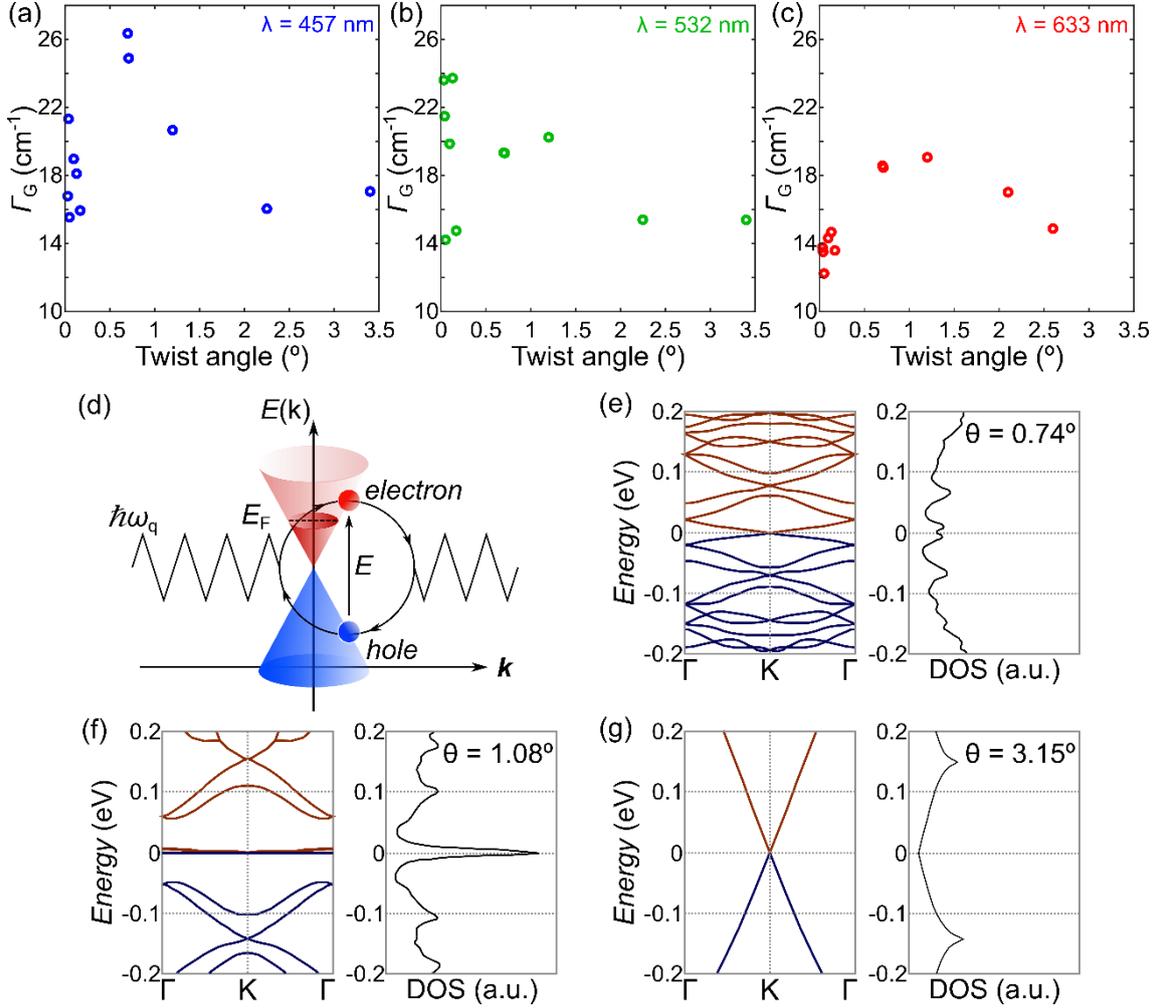

Fig. 5: G band full width at half maximum ($\Gamma_G$) dependence on the TBG twist angle for (a) 457 nm, (b) 532 nm, and (c) 633 nm (adapted from Gadelha et al. [1]) excitation laser wavelengths. The data reveal a maxima of $\Gamma_G$ close to $\theta_M$ for all measured laser lines, while the $\Gamma_G$ spread for $\theta \rightarrow 0°$ might be due to sample inhomogeneity; (d) Schematics of the Kohn anomaly process in graphene (adapted from Hasdeo et al. [62]); Band structure and DOS of TBG samples calculated through a tight-binding approach for (e) $\theta$ = 0.74°, (f) $\theta$ = 1.08° (magic angle), and (g) $\theta$ = 3.15° (adapted from Nguyen et al. [37]). The $\theta$ = 1.08° reveals a flat band at zero energy which is directly related to the spike in the DOS.

## 6. CONCLUSION

In summary, we presented an investigation of the Raman spectra of TBG with twist angles ranging from 0.03º to 3.40º, with three different laser lines, and our analysis focused on the Raman-active G and 2D bands. The spectra reveal two main $\theta$-dependent phenomena: (i) an increase of the G band width ($\Gamma_G$) close to the magic angle, which is understood based on electron-phonon interaction. The increase in $\Gamma_G$ is expected due to the larger density of states of samples close to the magic angle, increasing the electron-phonon interaction and lowering the phonon lifetime; (ii) a variation in the 2D band line shape, which indicates a general trend of an increased relative contribution from AA/SP regions as compared to the Bernal (AB/BA) regions to the Raman spectra of samples with twist angles smaller than the magic angle, according to previous TBG theoretical works [37]. The 2D bands were decomposed in two main contributions: a Bernal stacking spectrum and a pseudo-Voigt function, related to the AA/SP regions or to high angle TBG samples spectra, which further highlights the decrease in the Bernal stacking contributon to the overall spectra with increasing twist angle. We also proposed a geometric model to further analyze this trend, which became an interesting approach to estimate the soliton width towards small moiré pattern unit cells, found here to be in the $W_{SP} = (4 \pm 3)$ nm range. Therefore, our presented results describe the science of the Raman spectroscopy around the magic angle, and it brings up the importance of Raman spectroscopy as a fast and non-destructive technique in the characterization of twisted bilayer graphene devices, a powerful tool for identifying the twist angles of such samples. I.e., a proposed protocol to locate effective magic angle in disordered samples [66] is to raster scan the sample looking for regions with the highest G band linewidth and with small Bernal stacking spectrum contribution to the 2D band spectra.

## SUPPORTING INFORMATION

Additional experimental details on Raman hyperspectral images combined with sMIM images for the accurate determination of the twist-angles can be found in the supporting information.

## ACKNOWLEDGMENTS

This work was supported by CNPq (302775/2018-8 and INCT/ Nanomateriais de Carbono), CAPES (RELAII and 88881.198744/2018-01) and FAPEMIG, Brazil. K.W. and T.T. acknowledge support from the JSPS KAKENHI (Grant Numbers 19H05790, 20H00354 and 21H05233).

# SUPPORTING INFORMATION

1. SCANNING MICROWAVE IMPEDANCE MICROSCOPY OF THE TWISTED BILAYER GRAPHENE SAMPLES

Our twisted bilayer graphene (TBG) samples were raster scanned in a scanning Microwave Impedance Microscope to precisely locate homogeneous Moiré patterns regions, as shown in Fig. S1 – the insets show high resolution sMIM images of regions with different Moiré patterns which allow obtaining their Moiré lengths ($L_M$). Then, we calculate the relative twist-angle of each region through eq. 1 presented in the main text.

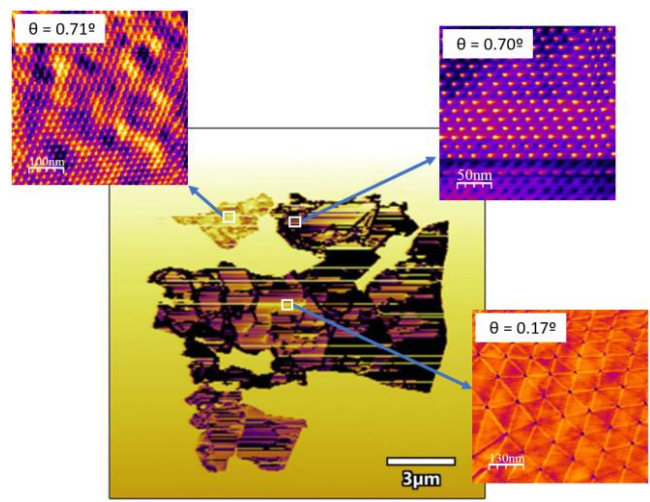

Fig. S1: sMIM image of a twisted graphene sample with the location of homogeneous Moiré patterns regions – the high-resolution images (see insets) reveal different Moiré lengths ($L_M$).

2. RAMAN HYPERSPECTRAL MAPS

To locate the homogeneous Moiré patterns regions in the Raman system, we acquired hyperspectral Raman maps of the TBGs (Fig. S2(a)) and combined them with the previously obtained sMIM image (Fig. S2(b)). Therefore, the sMIM images and the Raman maps were carefully aligned and superimposed (Fig. S2(c)) enabling the precise identification of the regions with well-defined twist-angle from which the Raman data were acquired.

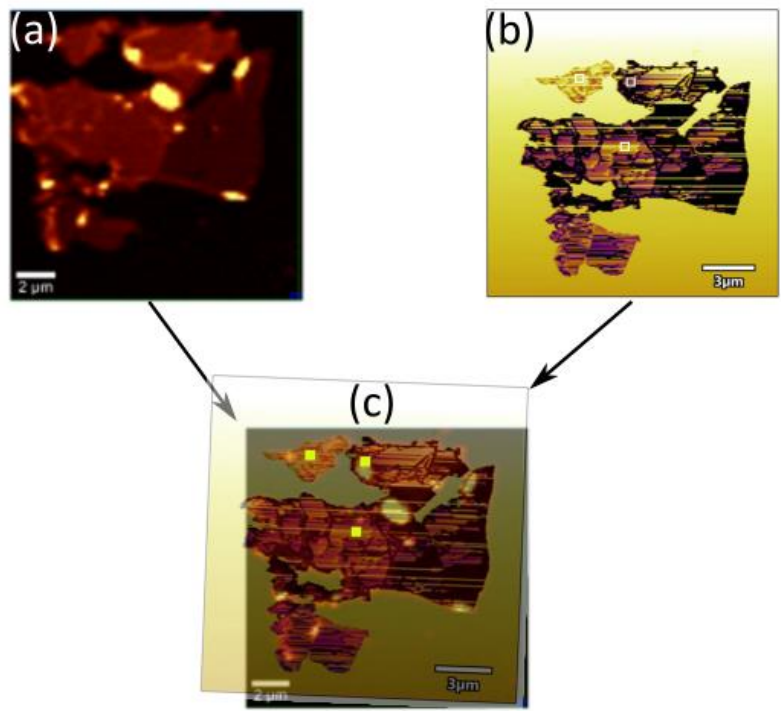

Fig. S2: (a) Hyperspectral Raman map of the G band of a TBG sample. (b) sMIM image of the same TBG sample presented in (a). (c) Superposition of the sMIM image over the Raman map enabling the precise identification of the homogeneous Moiré patterns regions prior to the acquisition of the Raman data.